\documentclass[twocolumn,a4paper,prc,floatfix,showpacs,preprintnumbers]{revtex4}
\usepackage{amssymb}
\usepackage{amsmath}
\usepackage[latin1]{inputenc}
\usepackage[dvips]{graphicx}

\begin{document}

\newcommand{\x}{{x_T}}
\newcommand{\ii}{{e_i}}
\newcommand{\jj}{{e_j}}
\newcommand{\y}{{y_T}}
\newcommand{\e}{{e_T}}
\newcommand{\kk}{{k_T}}
\newcommand{\p}{{p_T}}

\newcommand{\ptil}{\tilde{p}}
\newcommand{\ktil}{\tilde{k}}
\newcommand{\qtil}{\tilde{q}}

\newcommand{\q}{{q_T}}
\newcommand{\n}{{n_T}}
\newcommand{\oo}{{0_T}}
\newcommand{\bb}{{b_T}}

\newcommand{\ex}{e_x}
\newcommand{\ey}{e_y}
\newcommand{\ei}{e_i}

\newcommand{\nab}{\nabla_T^2}

\newcommand{\emu}{\! e_\mu \!}
\newcommand{\enu}{\! e_\nu \!}

\newcommand{\ud}{\, \mathrm{d}}
\newcommand{\uc}{{\mathrm{c}}}
\newcommand{\ul}{{\mathrm{L}}}
\newcommand{\intd}{\int \!}
\newcommand{\tr}{\, \mathrm{Tr} \, }
\newcommand{\R}{\mathrm{Re}}
\newcommand{\nc}{{N_\mathrm{c}}}
\newcommand{\nf}{{N_\mathrm{F}}}
\newcommand{\half}{\frac{1}{2}}
\newcommand{\hc}{\mathrm{\ h.c.\ }}
\newcommand{\nosum}[1]{\textrm{ (no sum over } #1 )}
\newcommand{\lqcd}{\Lambda_{\mathrm{QCD}}}
\newcommand{\as}{\alpha_{\mathrm{s}}}
\newcommand{\na}{\, :\!}
\newcommand{\nb}{\!: \,}
\newcommand{\cf}{C_\mathrm{F}}
\newcommand{\ca}{C_\mathrm{A}}
\newcommand{\df}{d_\mathrm{F}}
\newcommand{\da}{d_\mathrm{A}}
\newcommand{\nr}[1]{(\ref{#1})} 
\newcommand{\dadj}{D_{\mathrm{adj}}}
\newcommand{\ra}{R_A}

\newcommand{\gev}{\ \textrm{GeV}}
\newcommand{\fm}{\ \textrm{fm}}
\newcommand{\ls}{\Lambda_\mathrm{s}}

\newcommand{\init}{\textrm{init}}
\newcommand{\final}{\textrm{final}}

\title{Production of gluons in the classical field model for 
heavy ion collisions}
\author{T. Lappi}
\email[Email address: ]{tuomas.lappi@helsinki.fi}

\affiliation{
Department of Physical Sciences,
Theoretical Physics Division
}
\affiliation{
Helsinki Institute of Physics\\
P.O. Box 64,
FIN-00014 University of Helsinki,
Finland
}


\preprint{HIP-2003-14/TH}
\preprint{hep-ph/0303076}

\begin{abstract}
The initial stages of relativistic heavy ion collisions are studied
numerically in the framework of a 2+1 dimensional classical Yang-Mills
theory. We calculate the  energy and number densities and momentum
spectra of the produced 
gluons. The model is also applied to noncentral collisions. 
The numerical results are discussed in the light of RHIC
measurements of energy and multiplicity and other theoretical
calculations. Some problems of the present
approach are pointed out.
\end{abstract}

\maketitle

\section{Introduction}

In ultrarelativistic heavy ion collisions such as studied at RHIC and LHC
particle production in the central rapidity region is dominated by the gluonic
degrees of freedom in the nucleus. At sufficiently small $x$ the phase space 
density of these gluons is large, so one can try to treat them as a 
classical color field.
Let us first briefly review the model of \cite{mclv,kmclw,kvperus}
before turning to our results in Sec. \ref{sec:res} and their phenomenological
implications in Sec. \ref{sec:disc}. Our notation is essentially that of
\cite{mclv,kmclw,kvperus}.

The idea of \cite{mclv} was to model the high momentum degrees of
freedom of a nucleus as static random classical color
sources with a Gaussian probability distribution:
\begin{equation}\label{eq:korr}
\langle \rho^a(\x) \rho^b(\y) \rangle = g^2 \mu^2 \delta^{ab}\delta^2(\x-\y),
\end{equation}
where $\x$ and $\y$ are vectors in the transverse plane.
The classical color field generated by this source is then obtained from
the equations of motion
\begin{equation}\label{eq:ym}
[D_{\mu},F^{\mu \nu}] = J^{\nu}.
\end{equation}

This original formulation of the model is very simple, beyond the nuclear
radius $\ra$ it only depends on 
one dimensionful phenomenological parameter $\mu$ (related to $\ls$ introduced
in \cite{knv} by $\ls = g^2 \mu$) and the QCD coupling $g$ that does not run
in this classical approximation. One may, however, argue that 
the Gaussian probability distribution should be replaced by something
else, namely a solution of the ``JIMWLK'' renormalization group equation \cite{jimwlk}.

The McLerran-Venugopalan model \cite{mclv} describes the wave function of one nucleus. 
Nucleus-nucleus collisions were first studied in this framework in \cite{kmclw}.
The source current is taken to be
\begin{equation}\label{eq:twonucl}
 J^{\mu} =  \delta^{\mu +}\rho_{(1)}(\x)\delta(x^-) 
+ \delta^{\mu -}\rho_{(2)}(\x)\delta(x^+),
\end{equation}
where the color charge densities $\rho_{(m)}$ of the two nuclei
are independent.
In the region $x^- < 0$, $x^+ < 0$ which is causally connected to neither 
of the nuclei, the solution can be chosen as $A_\mu = 0$. In the regions
$x^- < 0$, $x^+ > 0$ and $x^- > 0$, $x^+ < 0$  
which are causally connected to only one of the nuclei the solutions are
``transverse pure gauges''
\begin{equation} \label{eq:ainit}
A_{(m)}^i = - \frac{i}{g} e^{ i\Lambda_{(m)}} \partial^i 
e^{ -i\Lambda_{(m)}}, \textrm{ with }
\nab \Lambda_{(m)} = -g \rho_{(m)}.
\end{equation}
The initial condition ($\tau = 0$) for the interesting region 
$x^- > 0$, $x^+ > 0$  is obtained by
matching the solutions on the light cone. This yields:
\begin{eqnarray}\label{eq:initcond}
A^i|_{\tau=0} &=& A^i_{(1)} + A^i_{(2)}, \\ \nonumber
A^\eta|_{\tau=0} &=& \frac{ig}{2}[A^i_{(1)},A^i_{(2)}].
\end{eqnarray}

Modeling the sources as delta functions on the light cone (Eq. \nr{eq:twonucl})
makes the initial conditions boost invariant. We shall also restrict ourselves
to strictly boost invariant field configurations.
This elimination of the longitudinal degrees of freedom makes the numerical solution
of the equations of motion easier, but is a serious limitation, especially
for studying thermalisation (see e.g. \cite{heinz2d}).

\section{(2+1)-dimensional classical Hamiltonian chromodynamics 
on the lattice}\label{sec:model}

The analytic solution of the equations of motion, Eq. \nr{eq:ym}, with the initial 
conditions, Eqs. \nr{eq:initcond}, is not known, but they can be studied numerically.
A lattice Hamiltonian formulation of the model was first
developed in \cite{kvperus}.

Assuming that the field configurations are boost-invariant reduces
the system to a 2+1-dimensional one. Choosing the Schwinger gauge $A_\tau = 0$ one
 can cast
the equations of motion into a Hamiltonian form. The lattice 
Hamiltonian is:
\begin{eqnarray}\label{eq:kogutsusskind}
aH &=& \sum_{\x} \Bigg\{ \frac{g^2 a}{\tau}\tr E^iE^i +
\frac{2\nc\tau}{g^2 a} \left( 1-\frac{1}{\nc}\R \ \tr U_\bot \right)
\nonumber \\
& &+\frac{\tau}{a} \tr \pi^2 +
\frac{a}{\tau} \sum_i
\tr \left(  \phi - \tilde{\phi}_i \right)^2 \Bigg\},
\end{eqnarray}
where $a$ is the lattice spacing and $E_i,U_i,\pi\textrm{ and }\phi$ are 
dimensionless lattice fields. The fields are matrices in color space, with
$E^i = {E_a}^i t_a$ etc. and the generators of the fundamental representation
normalised in the conventional way as $\tr t_a t_b = 1/2 \delta_{ab}$.
The first two terms are the transverse electric and magnetic fields, with
the transverse plaquette
\begin{equation}
U_\bot(\x) = U_x(\x)U_y(\x+\ex)U^\dag_x(\x+\ey)U^\dag_y(\x).
\end{equation}
The last two
terms are the kinetic energy and covariant derivative of the rapidity component of
the gauge field $\phi \equiv A_\eta = -\tau^2 A^\eta$,
which becomes an adjoint representation scalar with the assumption of boost invariance.
For the parallel transported scalar field we have used the notation 
\begin{equation}
\tilde{\phi}_i(\x) \equiv U_i(\x)\phi(\x+ \ii)U_i^\dag (\x).
\end{equation}
In the Hamiltonian, Eq. \nr{eq:kogutsusskind}, there is a residual invariance under 
gauge transformations depending only on the transverse coordinates. 
The Hamiltonian  equations of  motion are 

\begin{eqnarray}\label{eq:mo1}
\dot{U}_i &=& i \frac{g^2}{\tau}E^i U_i \nosum{i},
\\
\label{eq:mo2}
\dot{\phi} &=& \tau \pi,
\\
\label{eq:mo3}
\dot{E}^x &=& \frac{i \tau}{2 g^2} \left[U_{x,y}+U_{x,-y} - \hc \right]
- \textrm{trace} 
\nonumber \\
&& + \frac{i}{\tau} [\tilde{\phi}_x,\phi] 
\\
\nonumber
\dot{E}^y &=& \frac{i \tau}{2 g^2} \left[U_{y,x}+U_{y,-x} - \hc \right]
- \textrm{trace} 
\\ \nonumber 
&& + \frac{i}{\tau} [\tilde{\phi}_y,\phi] ,
\\
\label{eq:mo4}
\dot{\pi} &=& \frac{1}{\tau}\sum_i\left[ 
\tilde{\phi}_i + \tilde{\phi}_{-i} - 2\phi \right].
\end{eqnarray}

The Gauss law, conserved by the equations of motion, reads:
\begin{eqnarray}\label{eq:gauss5}
&\sum_i& 
\left[U^\dag _i(\x-\ii)E^i(\x-\ii)U_i(\x-\ii)  
-  E^i(\x)\right]  \nonumber \\
&-& i [\phi,\pi] = 0.
\end{eqnarray}

\begin{widetext}
On the lattice the initial conditions \nr{eq:initcond} become:
\begin{eqnarray}\label{eq:latinitcond}
0 &=& \tr \left[t_a \left(\left(U^{(1)}_i +U^{(2)}_i\right)
\left(1+U_i^\dag\right) - \hc \right) \right], \label{eq:init1} \\ 
E^i &=& 0, \label{eq:init2} \\
\phi &=& 0, \label{eq:init3} \\
\pi(\x) &=&  \frac{-i}{4g} \sum_{i} \bigg[
\left(U_i(\x) - 1\right)
\left(U_i^{\dag (2)}(\x)-U_i^{\dag (1)}(\x) \right)
\\ \nonumber & +& 
\left(U_i^{\dag}(\x-\ii) - 1\right)
\left(U_i^{(2)}(\x-\ii)-U_i^{(1)}(\x-\ii) \right) -\hc
\bigg], \label{eq:init4}
\end{eqnarray}
\end{widetext}
where $U^{(1,2)}$ in Eq. \nr{eq:init1} are the link matrices corresponding to 
the color fields of the two nuclei ($A_i^{(1,2)}$ in Eq. \nr{eq:ainit}) 
and the link matrix $U_i$ corresponding to the $\tau \ge 0 $ 
color field $A_i$ must be solved from
Eq. \nr{eq:init1}.

The model has three free parameters, the coupling $g$, the source density $\mu$
and the nuclear transverse area $\pi \ra^2$. In this work the lattice size 
is taken  to be $L^2 = N^2a^2 = \pi \ra^2$. This means that the field 
modes have an infrared cutoff of the order $1/\ra$, while
physically one would expect them 
to be cut off at a scale $\sim \lqcd$ by confinement physics not included in the
classical field model. So in order to be physically sensible our results 
should not depend on this infrared cutoff.

The values of the three parameters $g$, $\mu$ and $\pi \ra^2$
separately are needed when translating lattice units to physical units, but the
dimensionless parameter $g^2 \mu \ra$ controls the qualitative behavior of
the model; the weak coupling or weak field limit is reached for small values of 
this parameter (see also \cite{gavai}). To see this consider the system on a transverse 
lattice of spacing $a$. Now we have $\delta^2(\x) \sim 1/a^2$. Thus, from Eq.
\nr{eq:korr}, the charge density $\rho \sim g \mu / a$. The Green's
function of the operator $\nab$ in Eq. \nr{eq:ainit} is a logarithm, which
is parametrically constant. Thus $\Lambda(\x)$ is obtained by summing
contributions $ \sim g \cdot a^2 \cdot  g \mu / a$ from each of the $\sim \ra^2 / a^2$ 
cells (the area of a cell being $a^2$). Because the charges are distributed as
Gaussians with zero expectation value, their sum scales as a square root of 
the number of lattice sites, and we get 
$\Lambda(\x) \sim a g^2 \mu \cdot \sqrt{\ra^2 / a^2} \sim g^2 \mu \ra$.
Because of the exponentials of $\Lambda$ in Eq. \nr{eq:ainit} it is 
the magnitude of the dimensionless field $\Lambda$ that determines
the nonlinearity of the model.

The same argument can also be formulated in momentum space. The Poisson equation,
Eq. \nr{eq:ainit}, can be written as $\kk^2 \Lambda(\kk) = g \rho(\kk)$. One needs
a prescription to deal with the zero mode, the one chosen here is color neutrality
of the system as whole, $\rho(\kk=\oo) = 0 = \Lambda(\kk=\oo)$. Then the dominant
contribution comes from the smallest nonzero Fourier mode, $\kk \sim 1/\ra$.
In momentum space the correlator \nr{eq:korr} is
$\langle \rho(\kk) \rho(\p)\rangle \sim g^2 \mu^2 \delta^2(\kk+\p)$ with
$\delta^2(\kk) \sim \ra^2$. Thus 
$\Lambda(\kk) \sim g \ra^2 \rho(\kk) \sim g \ra^2 \cdot g \mu \ra$ and 
$\Lambda(\x) \sim \Lambda(\kk) / \ra^2 \sim g^2\mu\ra$.

In this Hamiltonian formalism the energy per unit rapidity in different field 
components is naturally the easiest and the 
most fundamental quantity to compute.
One can also measure equal time correlation functions of fields:
\begin{eqnarray}
&&\langle E^a_i(\kk,\tau)E^a_i(-\kk,\tau)\rangle, \\
&&\langle A^a_i(\kk, \tau)A^a_i(-\kk,\tau)\rangle,  \\
&&\langle \pi^a(\kk,\tau)\pi^a(-\kk,\tau)\rangle,\\
&&\langle \phi^a(\kk,\tau)\phi^a(-\kk,\tau)\rangle.
\end{eqnarray}
These correlation functions are 
not gauge invariant. One can, however, argue that in the Coulomb gauge
$\partial_i A_i = 0$ a physical meaning can be assigned to them 
(see also \cite{kvvanhat}).
Using equal time field correlation functions one can define a gluon 
number density $n(\kk)$, but the definition is not unique. The question of 
defining the number density is discussed in the following section.

\section{Particles in a classical field}

In a weakly interacting scalar theory it is easy to define a particle number
corresponding to a given classical field configuration. Take a free Hamiltonian
and Fourier transform it:
\begin{eqnarray}
H &=& \intd \ud^d x \left[ \half \pi^2(x) + \half (\nabla \phi)^2(x) +
\half m^2 \phi^2 \right]\\
&=& \intd \frac{\ud^d k}{(2\pi)^d}  \left[ \half |\pi(k)|^2 + 
\half \omega^2(k) |\phi(k)|^2 \right] \nonumber \\
&=& \intd \ud^d k \omega(k) n(k),
\end{eqnarray}
with the free dispersion relation $\omega^2(k)= k^2 + m^2$. 
Averaged in time the energy is distributed equally between the 
degrees of freedom:
\nopagebreak
\begin{equation}
\overline{\half |\pi(k)|^2} =
\overline{\half \omega^2(k) |\phi(k)|^2}
\end{equation}
so we can identify:
\begin{equation}
\overline{|\pi(k)|^2} = \omega(k) n(k), \quad
\overline{|\phi(k)|^2} =  \frac{n(k)}{ \omega(k)}.
\end{equation}

\begin{figure}[!htb]
\begin{center}
\includegraphics[width=0.5\textwidth]{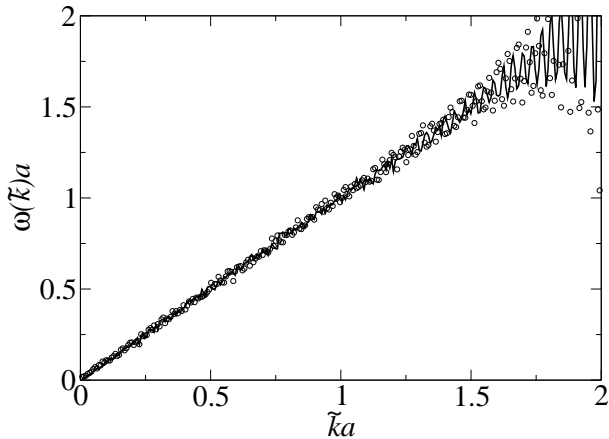}
\end{center}
\caption{The functions \nr{eq:omega}. The circles are $\omega(\ktil)$ determined
from the transverse fields $E^i$ and $A_i$, the solid line is
$\omega(\ktil)$ determined from $\pi$ and $\phi$. 
The maximum value of $\ktil a$ is $2\sqrt{2}$.
}\label{fig:omega}
\end{figure}

For an interacting theory one can \emph{define} the number distribution 
as follows:
\begin{equation}\label{eq:multi1}
n(k) = \sqrt{\overline{|\pi(k)|^2}  \  \overline{|\phi(k)|^2}}, \quad
\omega(k) = \sqrt{\frac{\overline{|\pi(k)|^2}}{\overline{ |\phi(k)|^2}}}.
\end{equation}
There is also another possibility, 
we can also \emph{assume} a dispersion relation 
\mbox{$\omega_{\mathrm{free}}(k) = \sqrt{m^2 + k^2}$} and define
\begin{equation}
n(k) = \frac{\overline{|\pi(k)|^2}}{\omega_{\mathrm{free}}(k)}.
\end{equation}
The latter approach is the one we take. Explicitly, for this particular 
theory described by the Hamiltonian, Eq. \nr{eq:kogutsusskind},
a 2-dimensional gauge field with an adjoint representation 
scalar field on the lattice, we define
\begin{equation}\label{eq:nspect1}
n(\kk) = \frac{2}{N^2} \frac{1}{\ktil}
 \left[ \frac{g^2}{2\tau} E^a_i(\kk)E^a_i(-\kk) 
+ \frac{\tau}{2} \pi^a(\kk) \pi^a(-\kk) \right],
\end{equation}
where 
\begin{equation}\label{eq:ktil}
\ktil^2 = \frac{4}{a^2} \left[\sin^2 \frac{a k_x}{2}
+ \sin^2 \frac{a k_y}{2} \right]
\end{equation} 
is the free, massless lattice
dispersion relation. We can then verify that our method is consistent
with the approach of Eq. \nr{eq:multi1} by looking at the correlation functions
\begin{equation}\label{eq:omega}
\frac{1}{\tau}\sqrt{\frac{ \langle E^a_i(\kk) E^a_i(-\kk)  \rangle}
{ \langle A^a_i(\kk) A^a_i(-\kk) \rangle}}
\quad \textrm{ and } \quad
\tau \sqrt{\frac{ \langle \pi^a(\kk) \pi^a(-\kk)  \rangle}
{ \langle \phi^a(\kk) \phi^a(-\kk) \rangle}}
\end{equation}
and verifying that they behave as $\omega(\ktil)\approx \ktil$ (see Fig.
\ref{fig:omega}). 

\section{Results}\label{sec:res}
\begin{figure}[!tbhp]
\begin{center}
\includegraphics[width=0.5\textwidth]{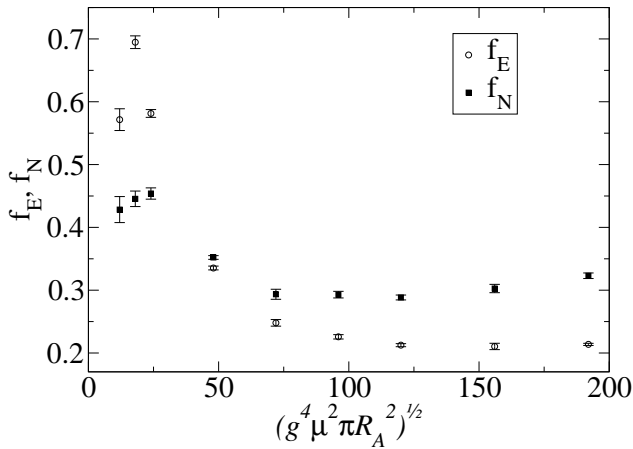}
\end{center}
\caption{The functions $f_E$ and $f_N$ as defined by Eqs. \nr{eq:deffe}, \nr{eq:deffn} 
vs. $\sqrt{g^4 \mu^2 \pi \ra^2}$. Computed on a $256^2$-lattice.
}
\label{fig:satur}
\end{figure}

\begin{figure}[!tbhp]
\begin{center}
\includegraphics[width=0.5\textwidth]{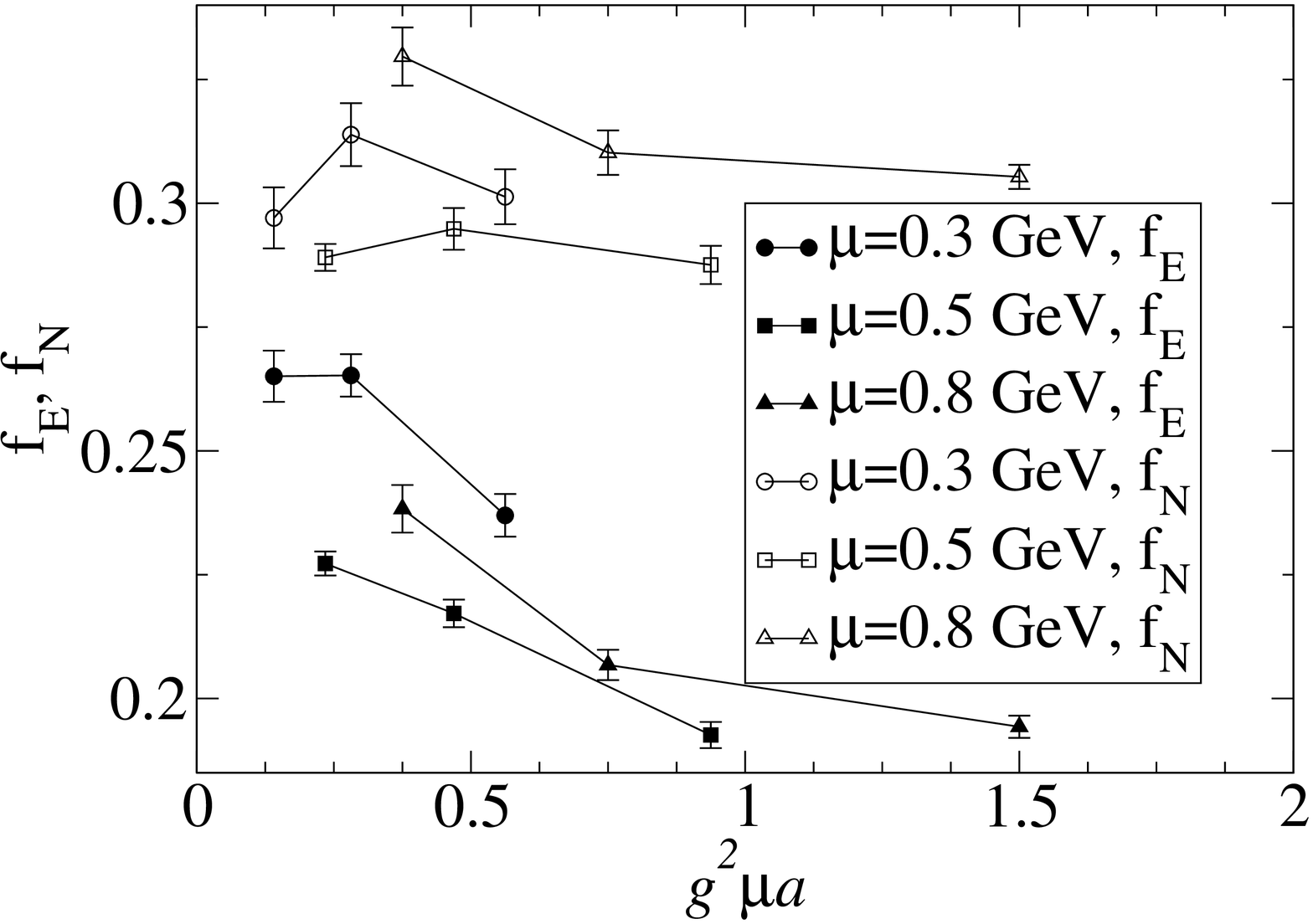}
\end{center}
\caption{
The functions $f_E$ and $f_N$ defined by Eqs. \nr{eq:deffe}, \nr{eq:deffn}
for constant $\sqrt{g^4 \mu^2 \pi \ra^2}$ and
with different lattice spacings. The horizontal axis is $g^2\mu a$, so the
continuum ($a \to 0$) limit is obtained by extrapolating each set of points
to the $g^2 \mu a =0$-axis on the left. 
}
\label{fig:contlim}
\end{figure}

\begin{figure}[!tbhp]
\begin{center}
\includegraphics[width=0.5\textwidth]{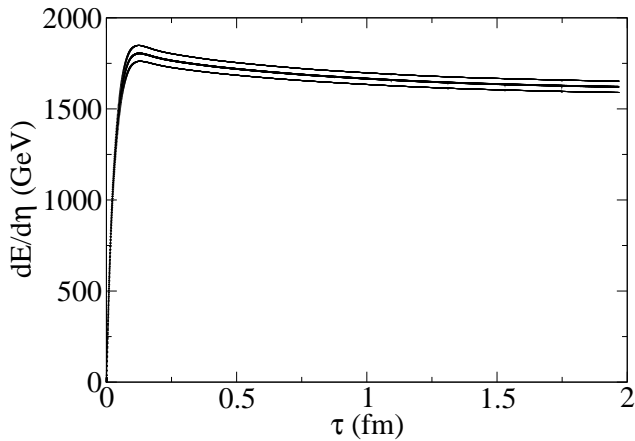}
\end{center}
\caption{Total energy per unit rapidity as a function of
time for $\mu = 0.5 \gev$. The three curves give an error estimate
from 5 trajectories on a $512^2$-lattice.}
\label{fig:toten}
\end{figure}

\begin{figure}[!tbhp]
\begin{center}
\includegraphics[width=0.5\textwidth]{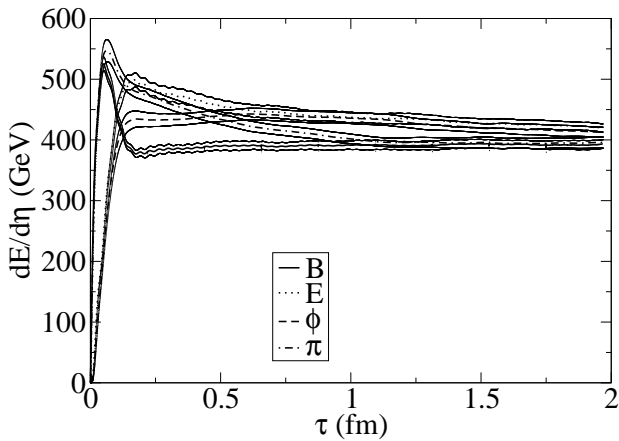}
\end{center}
\caption{
Energy in different field components from the same simulations as in Fig. 
\ref{fig:toten}.}
\label{fig:allen}
\end{figure}

To state our results in a form easily comparable with \cite{knv} 
let us define the same dimensionless quantities $f_N$ and
$f_E$ as follows:
\begin{eqnarray}\label{eq:deffe}
f_E &=& \frac{1}{g^4 \pi \ra^2 \mu^3}  \frac{\ud E_\init}{\ud \eta},\\
\label{eq:deffn}
f_N &=& \frac{1}{g^2 \pi \ra^2 \mu^2} \frac{\ud N_\init}{\ud \eta}.
\end{eqnarray}
As discussed in Sec. \ref{sec:model}
the quantities $f_E$ and $f_N$ are functions of only one dimensionless variable
$g^4 \pi \ra^2 \mu^2$. In the weak field limit, namely for 
$\sqrt{g^4 \pi \ra^2 \mu^2} \lesssim 50,$
$f_E$ and $f_N$ have a strong dependence
on $g^4 \pi \ra^2 \mu^2$. This signals a dependence on the infrared 
cutoff of the
theory. In the strong field limit, i.e. at large enough values of
$g^4 \pi \ra^2 \mu^2$, the nonlinearities of the infrared modes regulate
this infrared divergence
and  $f_E$ and $f_N$ become approximately independent of 
$g^4 \pi \ra^2 \mu^2$, as can be seen from Fig. \ref{fig:satur}.
Our results for the energy and multiplicity are summarized 
in Figs. \ref{fig:satur} and \ref{fig:contlim} and Table \ref{tab:results}.
The total energy as a function of time in different field components is plotted
in Fig. \ref{fig:toten} and the energy in the different field components
in Fig.  \ref{fig:allen}.

\begin{table}[!tbhp]
\begin{center}
\begin{tabular}{|c|c|c|c|}
\hline
$\sqrt{g^4 \mu^2 \pi \ra^2}$ & $\mu$ ($\!\!\gev$) & $f_E$ & $f_N$ \\
\hline
72 & 0.29 & 0.265 $\pm$ 0.005 &  0.297 $\pm$ 0.006\\
120 & 0.49 & 0.227 $\pm$ 0.003 & 0.289 $\pm$ 0.003 \\
192 & 0.78 & 0.238 $\pm$ 0.005 & 0.329 $\pm$ 0.006 \\
\hline
\end{tabular}
\end{center}
\caption{The values for $f_N$ and $f_E$ corresponding to the points nearest to the 
continuum limit in  Fig. \ref{fig:satur}. 
The value of $\mu$ in physical units is computed taking
$g=2$ and $\pi \ra^2 = 148 \fm^2$. The value of $f_E$ is obtained by fitting
the energy to a form $A + Be^{-\tau/\tau_0}$ and using the value $A$. 
The multiplicity is measured at a time $\tau = 5/\mu$, but its dependence on 
$\tau$ is very weak.
}\label{tab:results}
\end{table}

\begin{figure}[!tbhp]
\begin{center}
\includegraphics[width=0.5\textwidth]{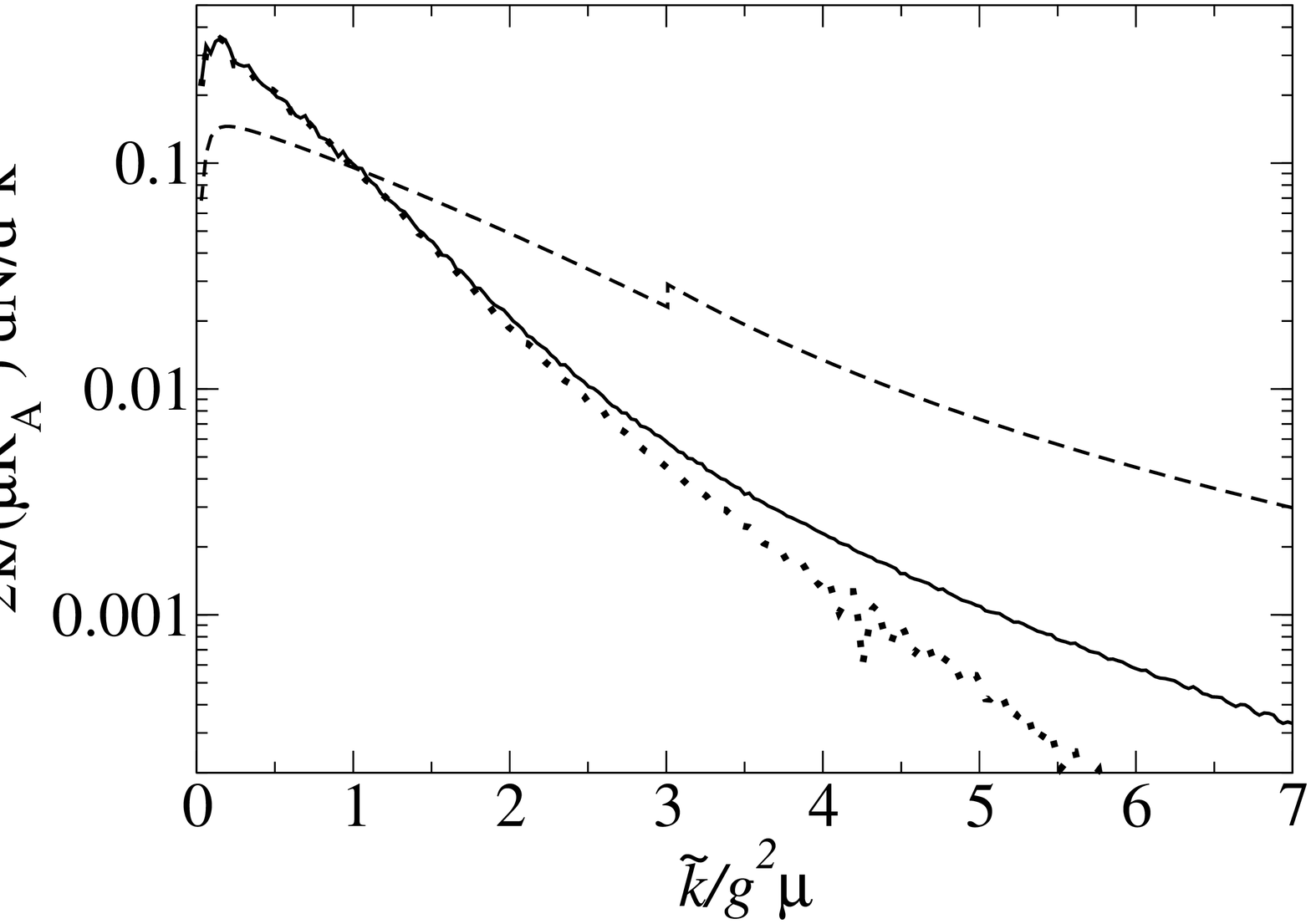}
\end{center}
\caption{
$\frac{2 \ktil}{\mu \ra^2}
\frac{\ud N}{\ud^2\kk}$ as a function of 
$\ktil/g^2\mu$ for $\sqrt{g^4\mu^2\pi \ra 2} = 120$.
The solid line is our result for a $512^2$-lattice,
the dotted line for a $256^2$-lattice and the dashed line a fit to
the numerical result of \cite{knv}.  The area under the curves (which is just
$f_N$ defined in Eq. \nr{eq:deffn}) is 
approximately the same (although the logarithmic scale makes this hard to see).
The dashed curve practically falls on top of the solid one if the vertical axis
is scaled by 2 and the horizontal by $1/2$ --- a signal of a difference in
the normalisation. }
\label{fig:normvert}
\end{figure}

Our result for $f_E$ is smaller than that of \cite{knv} by approximately a factor of
two. The function $\ud N / \ud^2\kk(\kk)$ 
we obtain is different although its integral over $\kk$-space, and thus $f_N$,
happens to be the same. This  difference is illustrated in 
Fig. \ref{fig:normvert}.

\begin{figure}[!tbhp]
\begin{center}
\includegraphics[width=0.5\textwidth]{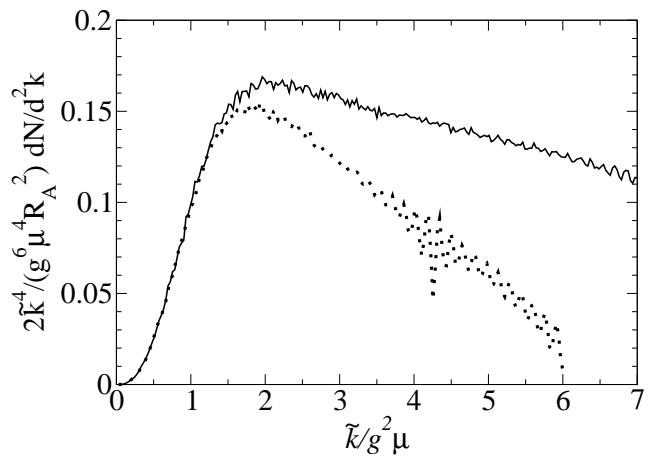}
\end{center}
\caption{
$\frac{2 \ktil^4}{g^6 \mu^4 \ra^2} \frac{\ud N}{\ud^2\kk}$ 
as a function of $\ktil/g^2\mu$  from the same simulations as Fig. \ref{fig:normvert}.
The solid line is our result for a $512^2$-lattice and
the dotted line for a $256^2$-lattice.}
\label{fig:k4hanta}
\end{figure}

\begin{figure}[!tbhp]
\begin{center}
\includegraphics[width=0.5\textwidth]{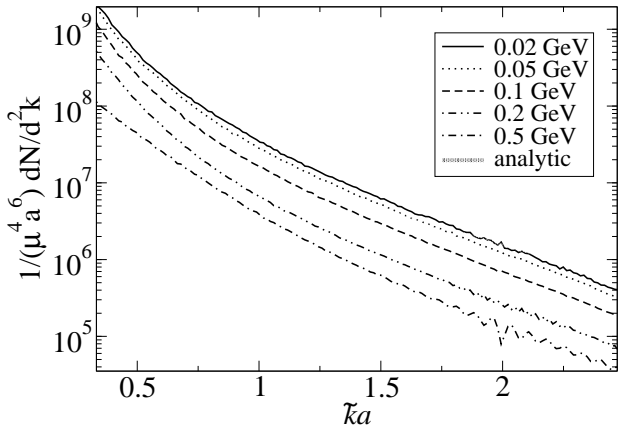}
\end{center}
\caption{
$\frac{1}{\mu^4 a^6}\frac{\ud N}{\ud^2 \kk}$ as a function of $\ktil a$ 
plotted for 
$\sqrt{g^4\mu^2\pi\ra^2} = 240 \mu/\!\!\gev$ with values of $\mu$ given
in the figure,  compared with
the analytical continuum result, Eq. \nr{eq:analcym}.} 
\label{fig:uusigtest}
\end{figure}

One can also derive a large $\kk$ analytic expression for the multiplicity in the 
classical field model \cite{kmclw} \cite{gyumcl} (see also \cite{dumitru}).
An expansion to the lowest nontrivial order in the field strength gives:
\begin{equation}\label{eq:analcym}
\frac{\ud N}{\ud \eta \ud^2 \kk} = 
\frac{\pi\ra^2}{(2\pi)^3}\frac{1}{\pi}\frac{\nc(\nc^2-1)g^6\mu^4}{\kk^4}
\ln \frac{\kk^2}{\Lambda^2},
\end{equation}
with $\Lambda$ some infrared cutoff. A useful check of the numerical computations
is that they should approach the analytic result in the weak field limit
of small $g^2\mu \ra$, although 
the uncertainty from the infrared divergence of the analytical result can be
numerically large.
Figure \ref{fig:k4hanta} shows that we do indeed observe a transition to a perturbative 
$1/\kk^4 \times$~logarithmic factors
-- behaviour around $\kk \gtrsim 2g^2\mu$, although in this region the shape of the 
spectrum is already severely modified by lattice effects, as can be seen comparing the
plots for the two lattice sizes.
But,  as seen in Fig. \ref{fig:uusigtest}, the overall normalisation of
our numerical result is far away from the analytical result at large
$\sqrt{g^4\mu^2\pi\ra^2}$ and approaches it only for
$\sqrt{g^4\mu^2\pi\ra^2}\lesssim 10$.
This would suggest that the weak field approximation used to obtain the
analytical result \nr{eq:analcym} is unsuitable for a quantitative 
understanding of this classical field model, whose justification lies, 
after all, in the argument of strong fields.

\begin{figure}[!tbhp]
\begin{center}
\includegraphics[width=0.5\textwidth]{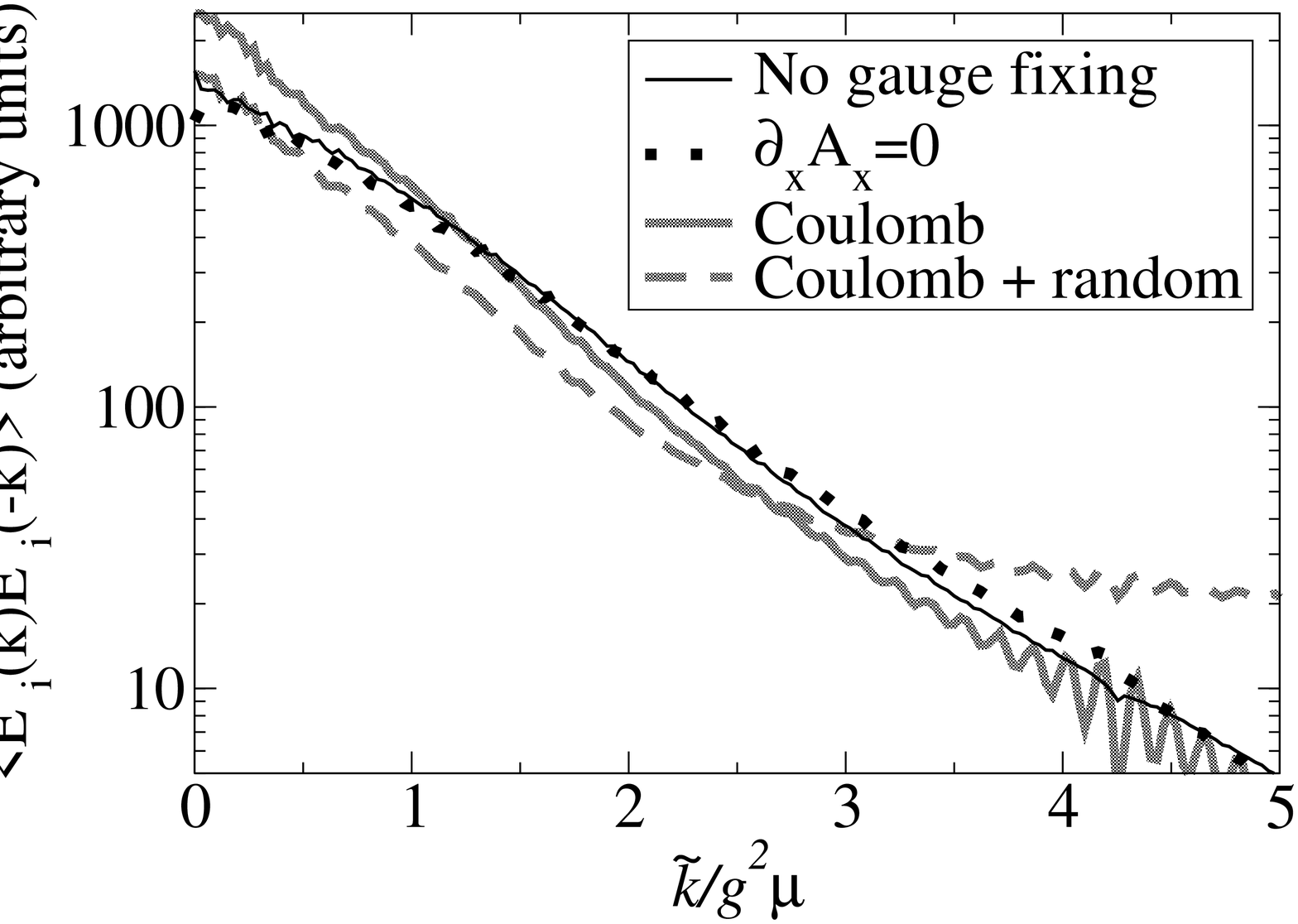}
\end{center}
\caption{
The correlator $\langle E^a_i(\kk) E^a_i(-\kk)\rangle$ in different gauges: 
the correlator resulting from the initial conditions and the equations of motion without
additional gauge fixing, in ``partial Coulomb'' $\partial_x A_x=0$ gauge, in 
the Coulomb gauge $\partial_iA_i = 0$ and in a gauge obtained by a
random gauge transformation of the Coulomb gauge field.}
\label{fig:spect}
\end{figure}

Our definition of the gluon number spectrum, Eq. \nr{eq:nspect1}, is based on equal
time correlators of fields. These correlators are gauge dependent, which
is a fundamental difficulty in defining a multiplicity of gluons for this 
classical field model.
We have studied this gauge dependence by using as an example the electric
field correlator $\langle E^a_i(\kk) E^a_i(-\kk)\rangle$, which is plotted
in Fig. \ref{fig:spect}.
Its gauge dependence is limited mainly by the constraint that the integral 
\begin{equation}
\int \ud^2 \kk  E^a_i(\kk) E^a_i(-\kk),
\end{equation}
which is proportional to the energy in the electric field, is gauge independent.

To determine the multiplicity using Eq. \nr{eq:nspect1} we take the fields resulting
from the initial conditions, Eqs. \nr{eq:latinitcond} and evolve them
in time according to the equations of motion, Eqs. \nr{eq:mo1}. The ``no gauge 
fixing''--curve in Fig. \ref{fig:spect} shows the 
$\langle E^a_i(\kk) E^a_i(-\kk)\rangle$-correlator obtained in this way. 
The fields are then  gauge transformed into the two dimensional Coulomb gauge
$\partial_iA_i=0$ to get the ``Coulomb gauge''-correlator, also plotted 
in Fig. \ref{fig:spect}. This is the one that is used to determine the multiplicity. 
In Fig. \ref{fig:spect}  we also plot the same correlator in two other gauges, 
$\partial_x A_x=0$ and a ``Coulomb + random'' gauge, which is obtained by 
taking a field configuration in the Coulomb gauge and
perfoming an independent random gauge transformation on each lattice site.  
In the latter the independent (Gaussian in this case)
transformations on each lattice site naturally enhance the high momentum parts
of the spectrum.

\begin{figure}[!tbhp]
\begin{center}
\includegraphics[width=0.5\textwidth]{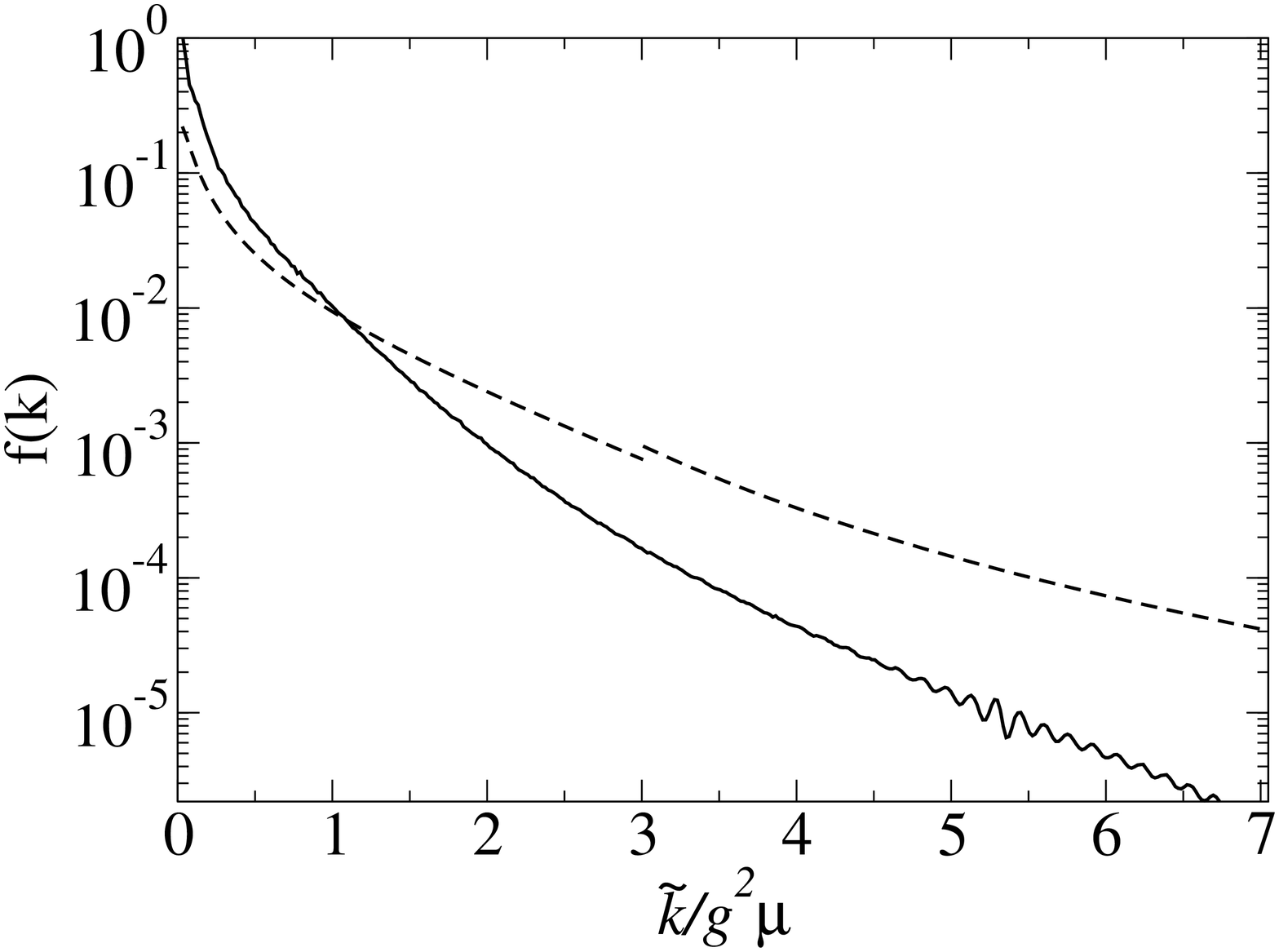}
\\
\includegraphics[width=0.5\textwidth]{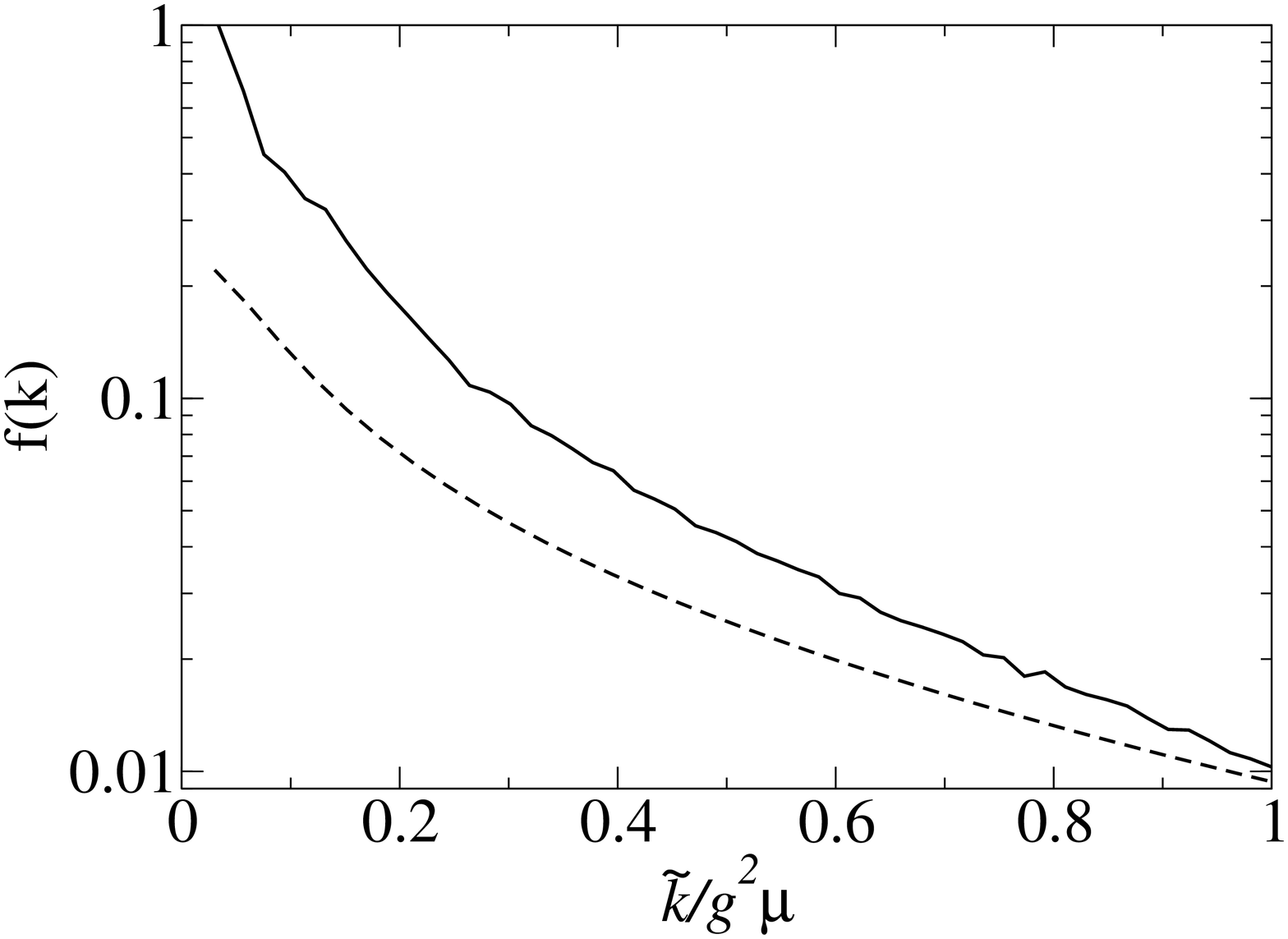}
\end{center}
\caption{The two dimensional phase space density 
$f(k) = \frac{1}{2 (\nc^2-1)}\frac{(2 \pi)^2}{\pi \ra^2}\frac{\ud N}{\ud^2 k}$
 as a function of $\ktil/g^2\mu$ for $\mu=0.5 \gev$ and $g=2$, 
i.e. $\sqrt{g^4\mu^2\pi\ra^2} = 120$.
The solid line is our result, the dashed line a fit to
the numerical result of \cite{knv}.}
\label{fig:phasesp}
\end{figure}

According to the discussion in Sec \ref{sec:disc}
the value of the parameter $\mu$ relevant to 
RHIC phenomenology would be $\mu = 0.5 \gev$ or $\ls = 2 \gev$. One can then ask 
whether this is indeed in the domain of validity of the present model, i.e. whether
the occupation numbers of gluons are high enough. To address this question
we plot in Fig. \ref{fig:phasesp} the two dimensional phase space density 
$f(\kk) = \frac{1}{2 (\nc^2-1)}\frac{(2 \pi)^2}{\pi \ra^2}\frac{\ud N}{\ud^2 \kk}$,
where the spin and color degeneracy has been divided out. It is of order one only 
up to momenta of a fraction of $g^2\mu$, meaning that the assumption of high
occupation numbers is only marginally satisfied.

Seeing that the results of \cite{knv,knvuusi} 
are in many aspects qualitatively similar to ours
and after a comparison of the numerical methods it seems that the difference
in our results concerning the energy and the number spectrum are simply due to 
a different normalisation of the SU(3) generators (compare Eq. \nr{eq:kogutsusskind}
and Eq. (A5) of \cite{knvuusi}). Any phenomenological 
discussion, such as the following, cannot be considered as an argument for 
the correctness of one or the other numerical result.

\section{Phenomenology}\label{sec:disc}
\subsection{What to expect}

To discuss the phenomenological implications of these results in the light
of RHIC experiments \cite{exp}
one must relate the calculated initial 
multiplicities and transverse energies to the observed quantities. There are
several scenarios that can be used to do this. Let us compare
different results with the assumption of early thermalisation and 
adiabatic expansion that has been successful in explaining particle yields and
elliptic flow. In this scenario the initial and final multiplicities, related
by entropy conservation, are approximately equal, and we take the total 
(charged and neutral) multiplicity per unit rapidity to be 
\begin{equation}
\frac{\ud N_\init}{\ud \eta} \approx \frac{\ud N_\final}{\ud \eta}  \approx 1000.
\end{equation}

The observed transverse energy is
\begin{equation}
\frac{\ud E_\final}{\ud \eta} \approx 600  \gev.
\end{equation}
The initial energy is larger than this, due to the expansion of the system.
In a freely streaming system the energy per unit rapidity is constant, 
whereas adiabatic longitudinal expansion makes it decrease as $\tau^{-1/3}$.
In \cite{ekrt1} the energy is found to be reduced by a factor of 3.5. Because
the calculation of \cite{ekrt1}
is done assuming a very early thermalization it gives an
upper bound to the reduction. This translates into a bound for the
initial transverse 
energy $\frac{\ud E_\init}{\ud \eta} \lesssim \  2100 \gev$. Thus a conservative 
estimate assuming ``parton-hadron duality'', be it from entropy conservation
or some other mechanism would be
\begin{equation}\label{eq:maxet}
\frac{\ud E_\init}{\ud \eta} 
\lesssim 2.1  \gev \frac{\ud N}{\ud \eta} .
\end{equation}

The final state saturation model of \cite{ekrt1}
is a pQCD-calculation supplemented by a sharp infrared 
cutoff determined from a simple geometrical final
state saturation argument. 
The result of the calculation is (Eq. (5) of \cite{ekrt1}):
\begin{equation}\label{eq:ekrt}
p_{\mathrm{sat}} \frac{\ud N_\init}{\ud \eta} =
0.288 \ \gev A^{1.050} (\sqrt{s})^{0.574},
\end{equation}
Setting $\ud N_\init/\ud \eta$ to $1000$ and taking $A=200$, 
$\sqrt{s} = 130 \gev$ gives 
$p_{\mathrm{sat}} = 1.23   \gev$. Then, from Eq. (7) of \cite{ekrt1}, we get
\begin{eqnarray}\label{eq:ekrten}
\frac{\ud E_\init}{\ud \eta}
&=& 1.43 p_{\mathrm{sat}} \frac{\ud N}{\ud \eta} \nonumber \\
&=& 1.76  \gev \frac{\ud N}{\ud \eta}.
\end{eqnarray}
Intrinsically such an unphysically sharp
infrared cutoff should produce too large an average energy per particle, because
there are no gluons with $p_T < p_{\mathrm{sat}}$ in the model.
The constant coefficient in front of \nr{eq:ekrt} is determined by the parton
distributions and is not fitted to match the RHIC data.
The result \nr{eq:ekrten} could thus be regarded as a
theoretical upper bound on the initial energy.

\subsection{Classical Yang-Mills result}

Let us take from Table \ref{tab:results} the result for $\mu \approx 0.5 \gev$, which
is the value of $\mu$ that gives approximately the right multiplicity. We get
\begin{equation}
\frac{\ud N_\init}{\ud \eta} = 0.29 g^2 \pi \ra^2 \mu^2 .
\end{equation}
This gives $\mu = 0.48 \gev$ or $\ls = 1.9 \gev$. Then the energy is
\begin{eqnarray}
\frac{\ud E_\init}{\ud \eta} &=& 0.23 g^4 \pi \ra^2 \mu^3 
\\
&=& 0.79 g^2 \mu  \frac{\ud N_\init}{\ud \eta}
\\
&=& 1.5 \gev \frac{\ud N_\init}{\ud \eta}.
\end{eqnarray}
This is well within the bound \nr{eq:maxet}.

The result of \cite{knv} is $f_N = 0.3$. Setting
$\ud N_\init / \ud \eta = 1000$ this gives us $\ls \ra = 65$. Taking
$\pi \ra^2 = 148 \fm^2$ this means $\ls = 1.87\gev$. For $f_E$ the result
in \cite{knv} is $f_E = 0.537$ for $\ls \ra = 25$ and  
$f_E = 0.497$ for $\ls \ra = 83.7$.
Taking the value $f_E = 0.5$ one gets
\begin{eqnarray}
\frac{\ud E_\init}{\ud \eta}
&=& 1.67 \ls  \frac{\ud N_\init}{\ud \eta}
\\
&=& 3.1 \gev \frac{\ud N_\init}{\ud \eta}.
\end{eqnarray}
Thus the average energy per particle is $3.1 \gev$, which is very hard to 
reconcile with the estimate \nr{eq:maxet} and forces one to either give up
the assumptions behind that estimate or conclude that RHIC energies
are not in the domain of validity of the classical field model. One can indeed
argue, as in \cite{knv}, that gluon number increasing processes lower
the average energy per particle in the subsequent evolution of the system.

\section{Finite nuclei}

It is easy to naively generalise the model to finite nuclei. The Gaussian
distribution of the random color charges is argued to arise from 
a sum of independent fluctuating charges. Thus it is the variance 
$\langle \rho^a(\x)\rho^b(\y) \rangle$ that should be proportional to the 
thickness of the nucleus:
\begin{equation}\label{eq:finite}
\langle \rho^a(\x)\rho^b(\y) \rangle =  
g^2 \mu^2 \delta_{\x,\y}\delta^{ab} T(\x-\x_0)
\end{equation}
with $T(\x) \sim \sqrt{\ra^2-\x^2}$
(or some more sophisticated thickness function).
Note that the normalisation of $\mu$ is different from the square 
nucleus case, here we fix it by the condition
\begin{equation}
\sum_{\x,\y} \langle \rho^a(\x)\rho^b(\y) \rangle = 
\delta^{ab} g^2 \mu^2 \pi \ra^2.
\end{equation}

One can then proceed as previously. But the problem one encounters is that
the colour fields generated by the sources have long Coulomb tails 
outside the nuclei. In two dimensions the initial colour fields
\nr{eq:ainit} decay only logarithmically away from the nuclei. Physically 
the colour fields should decay at distances $\sim 1/\lqcd$ due to
confinement physics not contained in this model.

The approach of \cite{knvv2} and \cite{knvuusi}, also advocated by \cite{lm},
is to directly address this question by
imposing colour neutrality of the sources at a length scale of the order
of a nucleon radius. But it is also possible that a proper inclusion 
of saturation effects in the probability distribution of the initial 
colour sources might cure this problem. Saturation does, after
all, suppress the very long wavelength modes responsible for the long tails.

\begin{figure}[!htb]
\begin{center}
\includegraphics[width=0.5\textwidth]{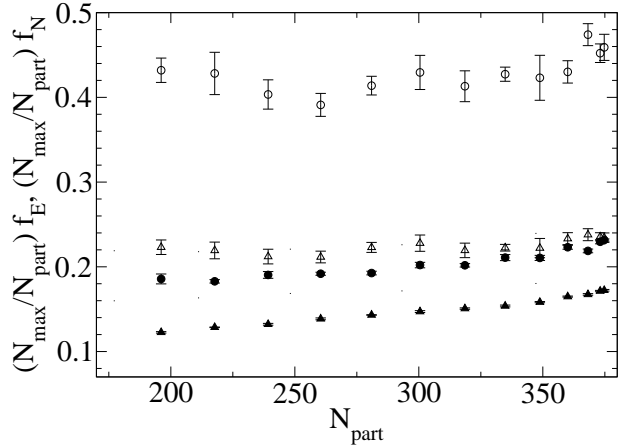}
\caption{The functions $\frac{N_\textrm{max}}{N_\textrm{part}}f_N$ (circles) and
$\frac{N_\textrm{max}}{N_\textrm{part}}f_E$ (triangles) vs. $N_\textrm{part}$. 
$N_\textrm{max} \approx 375$ is $N_\textrm{part}$ corresponding to impact parameter $b=0$.
The open symbols are results calculated with the original Gaussian weight function
and the filled symbols with the saturation ansatz \nr{eq:sat}.
The conversion from $b$ to $N_\textrm{part}$ from
\protect\cite{kimmo}.
}\label{fig:finite}
\end{center}
\end{figure}

Exploring the full implications of the ``JIMWLK'' renormalization group 
equation for heavy ion collisions is out of the scope of this work, but 
in the spirit of, e.g., \cite{iimcluusi} we have tried substituting the 
correlation function \nr{eq:finite} with the following procedure.
We take random variables $f^a(\x)$ distributed as:
\begin{equation}
\langle f^a(\x) f^b(\y) \rangle = \delta_{\x,\y}\delta^{ab} T(\x).
\end{equation}
The original McLerran-Venugopalan model, equation \nr{eq:korr} would be obtained with
the choice $\rho^a(\x) = g \mu f^a(\x)$. Now we Fourier transform and take
\begin{equation}\label{eq:sat}
\rho^a(\kk) =  g \mu \sqrt{\frac{\ktil^2}{\ktil^2 + g^4 \mu^2}} f^a(\kk).
\end{equation}
For $\ktil \gg g^2\mu$ this approaches the original McLerran-Venugopalan model, but 
for $\ktil \ll g^2 \mu$ the fluctuations are cut off as 
$\langle \rho^a(\kk) \rho^b(\kk)\rangle \sim \ktil^2$
Our results for the multiplicity and energy as a function of centrality are 
plotted in Fig. \ref{fig:finite}. All data points have been produced with the
same number, 10, of configurations, the larger errors seen using the original 
Gaussian weight function are a signal of its strong dependence on few infrared modes.
The discrepancy in the ratio $E/N$ between our results and those of 
\cite{knv,knvuusi} remains also in the finite nucleus case.

In \cite{knvuusi} it is said that ``our [using `color neutral' 
initial conditions] results may be \emph{quantitatively} similar to RG 
evolved predictions''. This can also be seen in Fig. 1
of \cite{knvuusi}, where the neutrality condition originally imposed
at the scale $\lqcd$ has an effect up to the scale $g^2\mu$, leading to
a modification of the Gaussian weight function that is very similar to ours.
It might thus turn out that at RHIC energies it is not yet possible to distinguish
effects from two physically very different phenomena, confinement and
saturation.

\section{Conclusions and outlook}

We have applied the classical field approach to heavy ion collisions and calculated
the energy and number densities and the spectra of the gluons produced in the
initial stages of the collisions. We have also extended the model to finite nuclei
and experimented with a crude saturation-inspired modification of the original model.
The gauge dependence of equal time correlators of the fields 
which makes it difficult to define a gluon number density has also been investigated.
A more practical difficulty in the model is that the phase space density of particles
at RHIC might not yet be large enough to justify its use, i.e.
the saturation scale might not be large enough compared to $\lqcd$.
For hard modes whose phase space density is small one does not even
expect a classical field approach to work, and the transition to
a pQCD regime should be understood better.

Further things that need to be investigated within this approach include
the incorporation of the ``JIMWLK'' renormalisation group equation
into the calculation. A better understanding of thermalisation, if
possible within the classical approach, might require extending
the study to a 3+1-dimensional model.
The ``best estimate'' in terms of physical postdictions for RHIC or predictions
for LHC phenomenology is not settled yet, but is hopefully 
converging.

\acknowledgements{
The author wishes to thank K. Kajantie for suggesting this topic and his advice,
F. Gelis, K. Rummukainen and K. Tuominen for numerous discussions and sharing their 
expertise and A. Krasnitz, Y. Nara and R. Venugopalan for discussions and correspondence.
This work was supported by the Finnish Cultural Foundation and the Academy of Finland 
(project 77744).}

\end{document}